\documentclass[12pt]{article}

\usepackage{graphicx}
\title{Strong decays and dipion transitions of $\Upsilon(5S)$
}
\author{  Yu.A.Simonov,
A.I.Veselov\footnote{ e-mail:veselov@itep.ru, fax:7(495)127 08 33}
\\State Research
Center\\Institute of Theoretical and Experimental Physics, \\
Moscow, 117218 Russia}
 \date{}
\newcommand{\beq}{\begin{eqnarray}}
 \newcommand{\eeq}{\end{eqnarray}}
\newcommand{\be}{\begin{equation}}
 \newcommand{\ee}{\end{equation}}

 \def\la{\mathrel{\mathpalette\fun <}}

\def\fun#1#2{\lower3.6pt\vbox{\baselineskip0pt\lineskip.9pt
\ialign{$\mathsurround=0pt#1\hfil ##\hfil$\crcr#2\crcr\sim\crcr}}}

\newcommand{{\SD}}{\rm SD}
\newcommand{{\Lc}}{\mathcal{L}}
\newcommand{{\Mc}}{\mathcal{M}}

\newcommand{\vep}{\mbox{\boldmath${\rm p}$}}
\newcommand{\veq}{\mbox{\boldmath${\rm q}$}}

\newcommand{\veK}{\mbox{\boldmath${\rm K}$}}

\newcommand{\vek}{\mbox{\boldmath${\rm k}$}}

\begin{document}
\maketitle
\begin{abstract}
Dipion transitions of  $\Upsilon (nS)$ with $n=5, n'=1,2,3$ are
studied using the Field Correlator Method, applied previously to
dipion transitions with $n=2,3,4$ The only two parameters of
effective Lagrangian were fixed in that earlier study, and total
widths $\Gamma_{\pi\pi} (5, n')$ as well as pionless decay widths
$\Gamma_{BB} (5S), \Gamma_{BB^*} (5S), \Gamma_{B^*B^*}(5S)$ and
$\Gamma_{KK} (5, n')$ were calculated and are in a reasonable
agreement with experiment. The experimental  $\pi\pi$ spectra for
$(5,1)$ and (5,2) transitions are  well reproduced  taking into
account  FSI in the  $\pi\pi$.
\end{abstract}
Pacs: 14.40.Nd; 13.25.Gv.

\section{Introduction}
 In a recent series of papers \cite{1}-\cite{3}, \cite{*} we have studied the $(n,
 n')$ bottomonium dipion transitions $\Upsilon (n) \to\Upsilon
 (n') \pi\pi$  and decays $\Upsilon (n) \to B \bar B, B \bar B\pi$ using effective Lagrangian derived in the framework
 of the  Field Correlator Method (FCM) \cite{4}. This Lagrangian,
 as was understood in \cite{3}, contains two effective masses,
 playing the role of decay vertices, $M_{\omega}$ for pionless
 $q\bar q$ pair creation, and $M_{br}$ for $q\bar q$ accompanied
 by one or two pions (kaons). It was found that $M_\omega$ is
 responsible for pionless decays of the type $\Upsilon (n) \to BB,
 B B^*, B^* B^*,$ while  $M_{br}$ enters into pionic decay
 transitions $\Upsilon(n) \to BB\pi$. These are the only free parameters of the method. It
 was shown in \cite{*}, that both pionless and dipion transition
 widths are reasonably well described by the method for $n=4,3,2$
 and $n'=1,2,3$  when theoretically sound values $M_\omega\sim
 \omega\approx 0.58$ GeV (average light quark energy in $B$) and $M_{br} \sim
 f_\pi \approx 93 $ MeV were used.

 The results of \cite{1}-\cite{3} allowed to describe the $\pi\pi$ spectrum
 in dipion $(n,n')$ transitions, for $n=2,3$ in \cite{1,2} and
 $n=2,3,4$  and $n'=1,2$ in \cite{3}. It was stressed in \cite{1}-\cite{3}, that the structure of
 the $(n,n')$ transition with  $BB, BB^*, B^*B^*$
 intermediate states contains two types of amplitudes: ``a" for
 consecutive  one-pion emission and ``b" for zero-pion --
 two-pion emission,  and the Adler Zero
 Requirement  (AZR) establishes connection between ``a'' and ``b''. In
 this way the long-standing problem of the theoretical description
 of all $(n,n')$ transition spectra, found in experiment
 \cite{5}-\cite{7} was approximately resolved. One should stress, however, that
 all $(n,n')$  dipion transitions in \cite{1}-\cite{3} with
 $n\leq4$ refer to the  subthreshold case, for $n=4$ the $BB$
 threshold is only 20 MeV below the $\Upsilon (4S)$ mass. For $\Upsilon (5S)$ the
 situation is different: all three channels $BB, BB^*, B^*B^*$  and three others with $B_s$ mesons are
 open and the corresponding imaginary parts are large due to large
 accessible energy. The final state $\pi\pi$ interaction is operative for the open channel
 amplitudes
 and one should calculate explicitly all terms in the amplitude, while AZR sets limits on the soft part of spectrum.

 The decays and transitions of $\Upsilon(5S)$ are a good check of
 our method, since no new parameters are involved, and the $5S$
 realistic wave function was accurately calculated \cite{8}. At
 the same time   the new experimental data on $5S$ decays \cite{9}
 present several questions for the theory:

 1) The  dipion widths $\Gamma_{\pi\pi}(5,1), \Gamma_{\pi\pi}
 (5,2), \Gamma_{\pi\pi}(5,3)$ are $\sim 1000$ times larger than
 the corresponding widths for $\Gamma_{\pi\pi} (nn')$ with
 $n=2,3,4.$

 2) The hierarchy of the widths $\Gamma_{BB} (5S)<
 \Gamma_{BB^*} (5S)< \Gamma_{B^*B^*}(5S)$occurs in experiment with $\Gamma_{tot} (5S)
 \sim 0(100$ MeV).

 3) Dikaon width of $\Upsilon (5S)$ is $\sim 1/10$ of the dipion
 width.

 4) The dipion spectra in (5,1), (5,2) transitions are not
 similar to spectra found for $n=2,3,4, $ showing a possible role
 of $\pi\pi$ FSI.

 It is a purpose of the present paper to study the $\Upsilon (5S)$
 decays and transitions using the same method as in
 \cite{1}-\cite{3} without introducing any new parameters. We
 shall give quantitative answers to questions 1) -4), finding a
 reasonable  order of magnitude agreement for all observables, however also a strong
 sensitivity to the properties of the $5S$ wave function.
  The paper is organized as follows. In section 2 general equations of the method from
  \cite{1}-\cite{3} are written for the case of $\Upsilon (5S)$. In
  section 3 pionless decay widths are computed and compared to
  experiment, whereas in section 4 total dipion and dikaon
  widths are discussed. The dipion spectra with and without
  $\pi\pi$ FSI factors are given in section 5. Main results are
  discussed in the concluding section together with a short
  summary and perspective.

  \section{General formalism for $\Upsilon(5S)$ decays and
  transitions}

  The amplitude of the dipion transition $(n,m)$ with pion momenta $\vek_1, \vek_2$ can be written
  according to \cite{3} as a sum of two terms, see Fig.1 (a),(b).

\vspace*{-4cm}
\unitlength 0.75mm 
 \linethickness{0.4pt}
\ifx\plotpoint\undefined\newsavebox{\plotpoint}\fi 
\begin{picture}(224.5,138.25)(0,0)
\put(51,51.25){\oval(56,24.5)[]} \put(134,51.75){\oval(50,25)[]}
\put(51.75,52.25){\oval(28.5,7)[]}
\put(221.25,134.5){\rule{3.25\unitlength}{3.75\unitlength}}
\put(135,52.25){\oval(27.5,7)[]} \put(37.75,52.75){\circle*{2.5}}
\put(65.5,52.5){\circle*{2.5}} \put(121.5,52.5){\circle*{2.5}}
\put(148.25,53.25){\circle*{.5}} \put(148.5,52.75){\circle*{2.55}}
\put(37.25,53.5){\circle*{1.12}} \put(38.5,52.75){\circle*{2.24}}
\put(37.43,53.18){\line(1,0){.95}}
\put(39.33,53.15){\line(1,0){.95}}
\put(41.23,53.11){\line(1,0){.95}}
\put(43.13,53.08){\line(1,0){.95}}
\put(45.03,53.05){\line(1,0){.95}}
\put(46.93,53.01){\line(1,0){.95}}
\put(48.83,52.98){\line(1,0){.95}}
\put(50.73,52.95){\line(1,0){.95}}
\put(51.68,52.93){\line(1,0){.5}}
\put(52.68,52.93){\line(0,1){0}}
\multiput(52.68,52.93)(-.0313,.0313){4}{\line(0,1){.0313}}
\put(64.93,53.18){\line(1,0){.972}}
\put(66.87,53.21){\line(1,0){.972}}
\put(68.82,53.24){\line(1,0){.972}}
\put(70.76,53.26){\line(1,0){.972}}
\put(72.71,53.29){\line(1,0){.972}}
\put(74.65,53.32){\line(1,0){.972}}
\put(76.6,53.35){\line(1,0){.972}}
\put(78.54,53.37){\line(1,0){.972}}
\put(80.49,53.4){\line(1,0){.972}}
\multiput(121.18,52.93)(.051961,.033333){15}{\line(1,0){.051961}}
\multiput(122.74,53.93)(.051961,.033333){15}{\line(1,0){.051961}}
\multiput(124.3,54.93)(.051961,.033333){15}{\line(1,0){.051961}}
\multiput(125.86,55.93)(.051961,.033333){15}{\line(1,0){.051961}}
\multiput(127.42,56.93)(.051961,.033333){15}{\line(1,0){.051961}}
\multiput(128.97,57.93)(.051961,.033333){15}{\line(1,0){.051961}}
\multiput(130.53,58.93)(.051961,.033333){15}{\line(1,0){.051961}}
\multiput(132.09,59.93)(.051961,.033333){15}{\line(1,0){.051961}}
\multiput(133.65,60.93)(.051961,.033333){15}{\line(1,0){.051961}}
\multiput(120.93,53.18)(.042484,-.033497){17}{\line(1,0){.042484}}
\multiput(122.37,52.04)(.042484,-.033497){17}{\line(1,0){.042484}}
\multiput(123.82,50.9)(.042484,-.033497){17}{\line(1,0){.042484}}
\multiput(125.26,49.76)(.042484,-.033497){17}{\line(1,0){.042484}}
\multiput(126.71,48.62)(.042484,-.033497){17}{\line(1,0){.042484}}
\multiput(128.15,47.49)(.042484,-.033497){17}{\line(1,0){.042484}}
\multiput(129.6,46.35)(.042484,-.033497){17}{\line(1,0){.042484}}
\multiput(131.04,45.21)(.042484,-.033497){17}{\line(1,0){.042484}}
\multiput(132.49,44.07)(.042484,-.033497){17}{\line(1,0){.042484}}
\put(42,25.5){\makebox(0,0)[cc]{Fig.1(a) Subsequent one-pion
emission.~~}} \put(122.75,25.25){\makebox(0,0)[cc]{Fig.1(b)
Two-pion emission.}}
\end{picture}
\vspace{-1cm}
$$ w_{nm}^{(\pi\pi)} (E) \equiv a-b= \frac{1}{N_c} \left\{ \sum_{n_2n_3} \int
\frac{d^3p}{(2\pi)^3}\frac{J_{nn_2n_3}^{(1)}(\vep,\vek_1)J^{*(1)}_{mn_2n_3}
(\vep,\vek_2)}{E-E_{n_2n_3}(\vep)-E_\pi(\vek_1)}+
(\vek_1\leftrightarrow \vek_2)\right.$$

$$-\sum_{n'_2n'_3}\int
\frac{d^3p}{(2\pi)^3}\frac{J_{nn'_2n'_3}^{(2)}(\vep,\vek_1,\vek_2)J_{mn'_2n'_3}^{*(0)}
(\vep)}{E-E_{n'_2n'_3}(\vep)-E(\vek_1,\vek_2)}-$$
\be\left.-\sum_{k^{\prime\prime}}\int
\frac{d^3p}{(2\pi)^3}\frac{J_{nn^{\prime\prime}_2n^{\prime\prime}_3}^{(0)}(\vep)J^{(2)*}_{mn^{\prime\prime}_2n^{\prime\prime}_3}
(\vep,\vek_1,\vek_2)}{E-E_{n^{\prime\prime}_2n^{\prime\prime}_3}(\vep)}\right\}\label{1}\ee

\bigskip

 \begin{center}

\hspace*{-3cm} \vspace{-1cm}

\includegraphics[width=8cm,keepaspectratio=true]{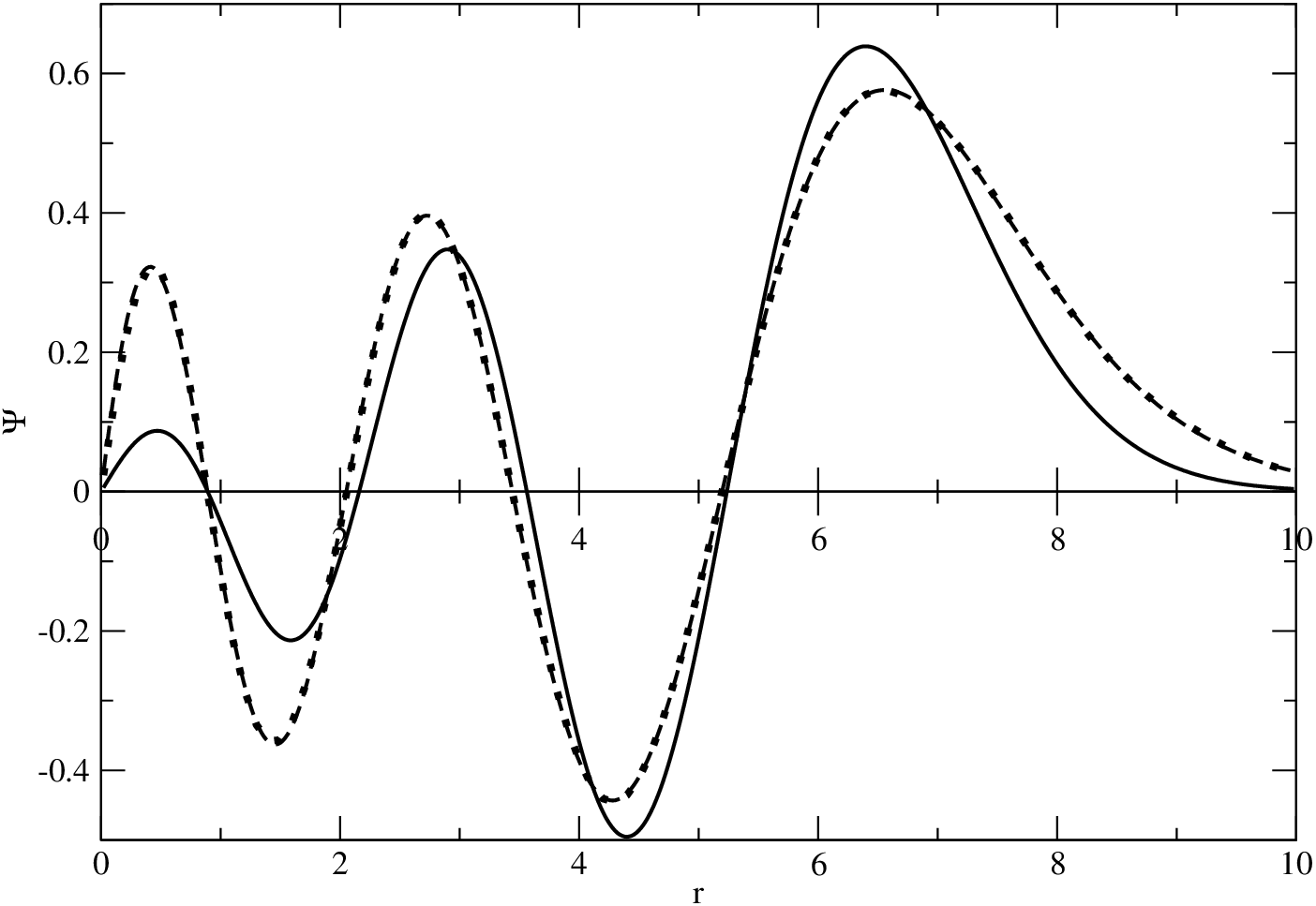}
\end{center}

 { Fig.2. Realistic w.f. of $\Upsilon (5S)$ (broken line),
the series of oscillator functions with
 $k_{\max}=15$ (dotted  line), $k_{\max}=5$ (solid). Note that the dotted curve is almost indistinguishable from the
 broken one.}


\bigskip

where $J^{(1)} (\vep, \vek), J^{(2)}(\vep, \vek_1, \vek_2)$ are
the overlap matrix elements between wave functions $\Psi(\veq)$ of
$\Upsilon(5S)$ and $\varphi(\veq_1)\varphi(\veq_2)$ of $B(B^*)$
mesons.

It is convenient to approximate $\Psi(\veq), \varphi(\veq)$ by a
series of oscillator wave functions; indeed in Fig. 2 we show the
quality of fitting of $\Psi(r)$ by  series of 5 and 15 terms. In
this case the dependence on $\vek_1, \vek_2$  as shown below
simplifies. For the pionless overlap matrix element one can write

\be J^{(0)}_{n,11} (\vep) =\int\frac{d^3\veq}{(2\pi)^3}\bar
y_{123} (p,q)\sum^{N_{\max}}_{k=1} c_k^{(n)} \varphi_k (\beta_1,
\veq +c\vep) \varphi^2_1 (\beta_2, \veq)=\frac{i p_i}{\omega}
e^{-\frac{\vep^2}{\Delta}} ~^{(1)}I_{n,11} (\vep)\label{2}\ee

Here  $c\approx 1$,  $c_k^{(n)}$ are $\chi^2$ fitting coefficients
and $\varphi_k$ --  oscillator functions  for $\Psi(q)$ and
$\varphi_1$ -- for $B, B^*$ mesons, and $\beta_1, \beta_2$ are
oscillator parameters for $\Upsilon(5S)$ and $B, B^*$ found from
fitting. The factor $\bar y_{123}$ defined in \cite{3} takes into
account the Dirac trace structure of the overlap vertex.

In a similar way one can define $J_n^{(1)}, J_n^{(2)}$ for one --
and two-pion emission integrals ($\veK=\vek_1+\vek_2$)

\be J^{(1)}_{n,11} (\vep, \vek) = e^{-\frac{\vep^2}{\Delta}-
\frac{\vek^2}{4\beta^2_2}} I_{n,11}(\vep)  \bar
y^{(\pi)}_{123}\label{3}\ee

\be J^{(2)}_{n,11} (\vep, \vek_1, \vek_2) =
e^{-\frac{\vep^2}{\Delta}- \frac{\veK^2}{4\beta^2_2}} ~^{(1)}
I_{n,11}(\vep)  \bar y^{(\pi\pi)}_{123} p_i.\label{4}\ee Here
$\bar y^{(\pi)}_{123}, \bar y^{(\pi\pi)}_{123}$ are defined by the
Dirac traces of the amplitudes and are given in \cite{3}. As a
result, the total amplitude is written as \be \Mc= \exp
\left(-\frac{\vek^2_1+\vek^2_2}{4\beta^2_2}\right) \left(\frac{
M_{br}}{f_\pi}\right)^2 \Mc_1 -\exp \left(
-\frac{\veK^2}{4\beta^2_2}\right) \frac{M_{br}
M_{\omega}}{f_\pi^2} \Mc_2.\label{5}\ee

 Here $\mathcal{M}_1 \sim a, \mathcal{M}_2 \sim b,$ explicit expressions for $\Mc_1, \Mc_2$ in terms of the integrals
of overlap matrix elements  $J^{(1)}, J^{(2)}, J^{(0)}$, as in
(\ref{1}), are given in \cite{3}, and here we only quote results
of numerical computations of $\Mc_1, \Mc_2$ for  (5,1), (5,2) and
(5,3) transitions. As  will be seen, both $\Mc_1$ and $\Mc_2$ do
not depend strongly  on $\cos\theta$ and $x$, so that the main
dependence of  $\Mc(x,\cos \theta)$ on arguments comes from two
exponential factors in (\ref{5}) (some exclusion is imaginary part
of $\Mc_1$, which is  peaked near $|\cos \theta |=1$).

The differential probability of dipion transition is given by
 \be \frac{dw_{\pi\pi} (n,n')}{dqd\cos\theta}
= C_0 \mu^2 \sqrt{x(1-x)}|\Mc |^2.\label{6}\ee where we introduced
variables $ q \equiv M_{\pi\pi}, q^2= (\omega_\pi(k_1) +\omega_\pi
(k_2))^2- (\vek_1 +\vek_2)^2$ $x= \frac{q^2-4m^2_\pi}{\mu^2}, ~~
\mu^2 \equiv (\Delta E)^2- 4 m^2_\pi;$ and numerical factor $C_0=
\frac{1}{32 \pi^3 N_c^2} = 1.12\cdot 10^{-4}$. Here $\Delta
E\equiv M(\Upsilon (nS)) - M (\Upsilon(n'S))$; explicit values of
$\mu $ and $\Delta E$ for ($5, n')$ transitions are the following
(in GeV); $ \Delta E (5,1)=1.4;~~ \mu(5,1)=1.37;~~ \Delta
E(5,2)=0.837,~~ \mu (5,2) =0.788; \Delta E (5,3) = 0.505, \mu
(5,3) =0.418$. Finally the total dipion width is given by \be
\Gamma_{\pi\pi} (n,n') =C_0 \mu^3 \int^1_0 dx
\sqrt{\frac{x(1-x)}{x+\frac{4m^2_\pi}{\mu^2}}} \int ^{+1}_{-1}
|\Mc (x, \cos \theta)|^2 \frac{d\cos \theta}{2}\label{7}\ee

\section{The $B$-meson decays of $\Upsilon(5S)$}

In this section we study the pionless decays of $\Upsilon (5S)$,
namely into $B\bar B, B\bar B^*+c.c., B^*\bar B^*, B_s\bar B_s,
B_s\bar B_s^*+c.c., B^*_s\bar B^*_s$ to which we ascribe numbers $
k=1,2,..6.$ The  corresponding formula for the width was derived
in \cite{3}, namely \be \Gamma(\Upsilon (nS) \to (B\bar B)^k)=
\left(\frac{M_\omega}{2\omega} \right)^2 \frac{p^3_kM_k}{6\pi N_c}
(Z_k)^2 |J^{BB}_{n}(p_k)|^2.\label{8}\ee

  $M_{k}$ is twice the reduced mass in channel $k$. The
corresponding   coefficients $Z_{k}$ account for spin and isospin
multiplicities and  (cf. similar coefficients in \cite{10}) are as
follows: \be Z^2_1=2Z^2_4=1,~~  Z^2_2=2Z^2_5=4,~~
Z^2_3=2Z^2_6=7.\label{9}\ee

Here $J^{BB}_n (p_k)$ are overlap matrix elements

\be \frac{p_i}{\omega} J^{BB}_{n} (\vep) =\int
\frac{d^3q}{(2\pi)^3} (q_i -\bar c p_i) \Psi^*_n (\vep+\veq)
\varphi^2_B(\veq)\label{10}\ee where $\bar
c=\frac{\omega}{2(\omega +\Omega)},$ and $\omega, \Omega$ are
average energies of light and heavy quarks in $B$ meson, computed
in \cite{11}, $\omega \approx 0.587$ GeV, $\Omega= 4.827$ GeV,
$\omega_s =0.639$ GeV, $\Omega_s = 4.83$ GeV, see Table 4 in
\cite{1}.

Expanding $\Psi_n ,\varphi_B$ in series of oscillator functions as
in \cite{3}, one obtains the form $J_n^{BB} (\vep) =
e^{-\frac{\vep^2}{\Delta} }~~^{(1)}I_{n11} (\vep)$, where
$~^{(1)}I_{n,11}(\vep)$ is a polynomial in $p^2$, $\Delta =
2\beta^2_1 +\beta^2_2$ and $\beta_1, \beta_2$ are oscillator
parameters for $\Upsilon(nS)$ and $B$ meson respectively, found
from the $\chi^2$ fitting procedure to the realistic wave function
calculated  in \cite{8}, and for $5S$ state and $B$ meson one
finds respectively $\beta_1=0.59$ GeV, $\beta_2=0.48$ GeV.

Denoting $\Gamma_k \equiv {\Gamma_{th} (\Upsilon (5S) \to channel
(k)})$, one has \be \left(\frac{2\omega}{M_\omega}\right)^2
\Gamma_k = 0.0177 p^3_k M_k Z^2_k |J_5(p_k)|^2\label{11}\ee
 where
$J_5 (p) =~^{(1)}I_{5,11}(p) e^{-\frac{p^2}{\Delta}}$, and
$~^{(1)}I_{5,11}$ is given in Eq.(\ref{2}).   Below in Table 1 the
computed values of $\Gamma_k$ for $ k=1,...6$ and with
$k_{max}=5$, i.e.five oscillator terms approximating wave function
of $\Upsilon(5S)$ are  given. Computing $^{(1)}I_{5,11}(p)$ for
different number of oscillator terms $k_{max}$, one can see, that
values of $I_{5,11}(t), ~~t=\frac{p^2}{\beta^2_0},~~ \beta_0
\approx 0.886$ GeV, in the interval $0.2\leq t \leq 2$ are
sensitive to $k_{max}$ and  vary around the value
$|I_{5,11}|\approx  1 $ GeV$^{3/2}$. We choose this value  to
estimate the variation of  $\Gamma_k$ and find that for the
dominant channel 3 the width changes by 6\%, while $\Gamma_4$ can
change by a factor of 10.

We now can compare our predicted theoretical values for $\Gamma_k$
with experimental data from \cite{12}.  First of all the total
width of $\Upsilon(5S)$ is known with 10\% accuracy,
$\Gamma_{tot}^{\exp}=110\pm 13$ MeV \cite{12}, and some relations
were established \cite{12} \be
\frac{\Gamma_1^{\exp}}{\Gamma_2^{\exp}}<0.92;
\frac{\Gamma_1^{\exp}}{\Gamma_3^{\exp}}<0.3;
\frac{\Gamma_2^{\exp}}{\Gamma_3^{\exp}}=0.324.\label{12}\ee

For channels with $B_s, B_s^*$ one has \cite{12}\be
\frac{\Gamma^{\exp}_4+\Gamma^{\exp}_5 +
\Gamma^{\exp}_6}{\Gamma_{tot}}=0.16\pm 0.02\pm 0.058\label{13}\ee
and also \be  \frac{\Gamma_4^{\exp}}{\Gamma_6^{\exp}}<0.16;
\frac{\Gamma_5^{\exp}}{\Gamma_6^{\exp}}<0.16.\label{14} \ee

Calculating $ \Gamma_{tot}$ from Table 1, one has $\Gamma_{tot}
\simeq \left( \frac{M_\omega}{2\omega}\right)^2 160$ MeV and
choosing $\left(\frac{M_\omega}{2\omega} \right)^2=0.6$  one can
approximately reproduce  the decay $\Upsilon(4S) \to B\bar B$
$\Gamma_{tot} \simeq  26$ MeV $vs~\Gamma_{\exp} =20.5\pm 2.5$ MeV
(see \cite{3}),while for $\Gamma_{tot} (5S)$  one has
  $\Gamma_{tot} =113$ MeV,
which is not far from the experimental value $\Gamma^{exp}_{tot}=
(110\pm 13)$ MeV. However for more accurate calculation of
$\Gamma_k$ one needs better knowledge of the wave function.

Comparing partial widths from the Table 1 with  experimental
limits (\ref{12})-(\ref{14}), one can see, that all inequalities
except the last right ones  in (\ref{12}) and (\ref{14}) are
satisfied by our theoretical values, however more work on
theoretical side (explicit form of $5S$ wave function) and in
experiment is needed.

\section{Dipion and dikaon transitions of $\Upsilon (5S)$}

In this section we discuss dipion spectra and angular
distributions  for  the transitions (5,1), (5,2) and (5,3), as
well as total dipion and dikaon widths, given by Eq. (\ref{11}).
The differential probability $\frac{dw_{\pi\pi}}{dq d\cos\theta}$
is given in (\ref{6}), and integrating over $dx$ or over $d\cos
\theta$ we obtain one-dimensional spectrum \be \frac{dw}{dq}
=C_0\mu^2\sqrt{x(1-x)}\int^{+1}_{-1} |\Mc|^2
d\cos\theta\label{15}\ee and angular distribution
\be\frac{dw}{d\cos \theta} =\frac12 C_0 \mu^3\int^1_0 dx
\sqrt{\frac{x(1-x)}{x+\frac{4m^2_\pi}{\mu^2}}} |\Mc(x,
\cos\theta)|^2\label{16}\ee

The values of $\Mc$, Eq.(5), were calculated using
$\frac{M_\omega}{M_{br}}=6$ and for $\Mc_1, \Mc_2$ the same
equations (23-25) from \cite{3} were used as for $\Upsilon (nS)$
transitions with $n\leq 4 $.

 At this point we impose on the amplitude  $\mathcal{M}$ the soft
 pion property, and use the AZR to rewrite Eq.(\ref{5}) in the
 form
 \be
 \mathcal{M}=\bar \mathcal{M}(\exp_1-\exp_2 f(q)),\label{0}\ee
 where  $\exp_1$ and $\exp_2$ refer to the exponential factors in (\ref{5}) and
 the factor $f(q)$, later used for the FSI effects, obeys the
 condition  $f(q^2=m^2_\pi) =1$. Normalizing $\bar \mathcal{M}$ to
 $\mathcal{M}_2$, so that $\bar \mathcal{M}=
 \frac{{M}_{br} M_\omega}{f^2_\pi} M_2$, one can insert (\ref{0}) in
 (\ref{15}) to  obtain $\Gamma_{\pi\pi}$. The corresponding values
 without FSI, i.e.  for $f(q) \equiv 1$  are given in Table 2,
 upper line, and called the model 1.

For the dikaon (5,1) transition one can in first approximation
neglect the change of $m_\pi$ to $m_K$ in matrix element (5), and
take  it into account in phase space, also remembering   that
$\Mc$ is $O\left(\frac{1}{f^2_\pi}\right)$, which should be
replaced by $O\left(\frac{1}{f^2_K}\right)$. In the total width
$\Gamma_{KK}$ (5,1) one can write similarly to (\ref{7})
\be\Gamma_{KK} (5,1) = C_0 \mu^3_K\int^1_0 dx \sqrt{\frac{
x(1-x)}{x+ \frac{4 m^2_K}{\mu^2_K}}}\int^{+1}_{-1} \frac{d\cos
\theta}{2} |\Mc_k|^2.\label{17a}\ee

Here $\mu^2_K =(\Delta E)^2-4 m^2_K=0.985$ GeV$^2$, $ \mu_K=0.992
$ GeV.

As a result, approximating the ratio of integrals over $dx$ as
$1/2$, one obtains \be \frac{\Gamma_{KK}
(5,1)}{\Gamma_{\pi\pi}(5,1)} =\frac12 \left(
\frac{\mu_K}{\mu}\right)^3 \left( \frac{f_\pi}{f_k}\right)^4 =
0.194 \left( \frac{f_\pi}{f_K}\right)^4 = 0.092 \approx
1/10,\label{18a}\ee where we have used $f_\pi =93$ MeV, $ f_K=112$
MeV \cite{12}.

Correspondingly one obtains the last column in Table 2 from the
second one, using (\ref{18a}).

\section{ Final state interaction in $(5, n')$ transitions}

One  of the important new features of $(5S)$ (and higher states
like $(6S)$ )  transitions is that a large phase space  is
available where  both $\sigma$ and $f_0$ resonances can be seen.
In (4,1) transitions  $f_0$ is at the edge of phase space  while
$\sigma$ in most transitions lies near the region $x=\eta$, where
amplitudes vanish and therefore no strong FSI effects are visible
in $(n,n')$ for $n\leq 4$.

In (5,1), (5,2) transitions the situation is different and e.g. in
the  (5,1) transition the $f_0$ resonance is well inside the
available $q$ region.

At this point it is necessary to stress that the  FSI acts
differently on one-pion (''a'' or $\Mc_1$) amplitude and two-pion
("b" or $\Mc_2$) amplitude. Namely, for the case of $\Mc_1$, where
two pions are emitted from two points separated by distance $L\sim
1/\Gamma,~~ \Gamma\la 0 (10 $ MeV), the $\pi\pi$ interaction of
range $r_0 \la 0.6\div 0.8$ fm is damped by a factor  of the order
of $r_0/L\sim O(1/10)$. E.g. in the  FSI description in
\cite{13}-\cite{15}, the relative weight of $\pi\pi$ amplitudes
with and without FSI was estimated as $\sim (1/7)$.

Completely different situation occurs in $b,(\Mc_2)$, where a pair
of $s$-wave pions with $I=0$ is emitted from a point (or, rather,
a region of the order of $\lambda\sim 0.1$ fm , $\lambda$ -
gluonic correlation length of QCD vacuum ). Here FSI is obligatory
and is given by the Omn$\grave{\rm e}$s-Muskhelishvili solution $
f(q) = \frac{P(q^2)}{D(q^2)} ;  $ with $P(q^2)$ -- a  polynomial
normalizing $f(q^2)$ at some point: we shall use normalization
$f(q^2=(2m_\pi)^2)=1$;
 a very close result is obtained for the Adler zero normalization
 $f(q=m_\pi)=1.$ Hence one can write $f(q^2)$ as follows (cf the corresponding factors in \cite{13,14}).
 \be
 f(q) =\alpha f_\sigma (q) +\beta  f_{f_0}(q)\label{00}\ee

 \be f_i(q^2) = \frac{D_i(q^2=4m^2_\pi)}{D_i(q^2)};~~ D_i(q^2) = \exp \left(
 -\frac{q^2}{\pi} \int^\infty_{4m^2_\pi} \frac{dq'^{2}
 \delta_i(q'^2)}{q'^2 (q'^2-q^2)}\right), ~~i=\sigma, f_0\label{17}\ee
 and $\delta_i (q^2)$ is the $\pi\pi$ phase due to the $i$-th
 resonance.

 In the simplest approximation one can write

\be f_\sigma (q) =  \left [
\frac{m^2_\sigma-m^2_\pi)^2+\gamma^2_\sigma}{(m_\sigma^2-q^2)^2+\gamma^2_\sigma}\right]^{1/2},~~
f_{f_0}(q) =\left [
\frac{(m^2_{f_0}-m^2_\pi)^2+\gamma^2_{f_0}}{(m_{f_0}^2-q^2)^2+\gamma^2_{f_0}}\right]^{1/2}sign
(m_{f_0}-q) .\label{20}\ee

 The factors,
corresponding to the resonances yield peaks, in (\ref{20}) the
$\sigma$ peak is a wide structure, while $f_0$ produces a sharp
peak near 1 GeV. Another feature of $f_{f_0}(q)$, Eq. (\ref{20}),
is that it changes sign just above position of $f_0$ due to the
jump of $\delta(q^2)$ nearly  equal to $\pi$, near $q=1$ GeV,
\cite{13,14}.

We have fitted the experimental (5,1) and (5,2) $\pi\pi$ spectra
using the form (\ref{0}) with $f(q)$ given in (\ref{00}) and
obtain the following values of parameters: $m_\sigma =0.5$ GeV,
$m_{f_0} =1.15$ GeV, $\gamma_\sigma=0.35$ GeV, $\gamma_{f_0}=0.1$
GeV; $\alpha =1,\beta=0.01$. We call this fit the model 2.

The resulting curves  (solid lines) are given in Figs.3 and 4 for
(5,1) and  in Figs. 5 and 6 for the (5,2) cases, together with the
curves for  the model 1 ($f\equiv 1, $ no FSI), shown by broken
lines. Note, that in Figs. 3-6  theoretical curves were fitted to
the experimental width  $\Gamma^{\exp}_{\pi\pi}$, which means that
$M_{br}/f_\pi$ were varied in the interval $1\div 0.75$.

\section{Results and discussion}

We start with the $BB$  widths of $\Upsilon (5S)$ given in Table
1. It is clear that the  values   $\Gamma_k$ give only a rough
estimate and actual values $\Gamma_k$ depends strongly on the
behaviour of the  $\Upsilon (5S)$ wave function. This is certainly
true for the Eq.(\ref{8}), derived for the wave function in the
one-channel approximation. In the next orders, given by the
equation
$$ \det \left( ( E-E^{(0)}_n)\delta_{nm} - w_{nm} (E)\right) =0,
$$ this sensitivity should be weaker, since the wave function
becomes complex and does not have zeros. Hence one might hope that
the  values $\Gamma_k$ yield the correct order of magnitude for
all channels $k=1,..6$, with the value
$\left(\frac{M_\omega}{2\omega}\right)^2\approx 1/2$ as deduced
from  $\Gamma_{tot}(\Upsilon(4S))$. Comparing $\bar \Gamma_k$ with
the widths $\Gamma_k$ obtained for the $5S$  wave function
approximated by 5 oscillator  functions, one finds a reasonable
agreement in magnitude , except for $\Gamma_4$ which is small due
to nearby zero of $J_5(p)$.

Coming now to the total dipion widths in Table 2, one can notice,
that our general expression (\ref{5}), without  FSI, yields
reasonable order of magnitude for $\Gamma_{\pi\pi}$ and
$\Gamma_{KK}$ if $\left(\frac{M_{br}}{f_\pi}\right)\approx 1$.
Here again strong  dependence on the  $\Upsilon(5S)$ wave function
persists and results for $k_{\max} =5$ and $ k_{\max}=15$ differ
several times.
 In view of this  it is not surprising that in Table 2 theoretical
 widths for (5,1) and (5,2)  dipion transitions have a hierarchy
 different from that of experimental widths; however the smallness
 of $\Gamma_{th} (5,3)$ is well explained by a small phase space
 factor $\mu^3: \mu^3(5,3)/\mu^3(5,1)\approx 2.8\cdot 10^{-2}$ and
 it is not clear, why $\Gamma_{exp} (5,3) \approx \Gamma_{exp}
 (5,1)$.

Similar results for $\Gamma_{\pi\pi}, \Gamma_{KK}$ are obtained
 when both FSI and AZR are taken into account.

Turning to the $\pi\pi$ spectra, one observes that the spectra
without FSI  (model 1) in Figs.3,5 have less structure  in
contrast to the experimental data \cite{9}, where  peaks in
spectra at $q=0.6$ GeV for (5,2)  and at  $q\cong 1.2$ GeV for
(5,1) are clearly seen and  strong $\cos \theta$ dependence is
observed for the (5,2) transition,

The situation is much better for the FSI-AZR approximation  (model
2)  in Figs. 3,5  where the $\sigma$ and $f_0$ peaks are seen  in
(5,2) and (5,1) cases, and also the experimental $U$- form of the
$\cos \theta$ distribution is produced in the   (5,2) transition.
However  the much weaker experimental  $\cos \theta$ dependence,
Fig. 4 for the (5,1) case is better reproduced in the model 1.

As a whole, it seems, that the spectrum, especially its lower
enhancement at $q\approx 0.4$ GeV in both (5,1) and (5,2)
transitions, can be well described by the AZR+ FSI form, where the
lower peak at $q\approx 0.4$ GeV is due to cancellation of two
terms in (\ref{0}), i.e. mainly due to AZR.

Summarizing, we have used the theory developed in previous papers
\cite{1}-\cite{3} and applied in \cite{3} to the subthreshold
transitions $(n, n'), n\leq 4$. This theory does not contain free
parameters, the only ones $M_\omega$ and $M_{br}$ are defined
previously in \cite{3}.

Exploiting   this theory, we have calculated six $BB$-type widths
of $\Upsilon(5S),$ $ \Gamma_k, k=1, ...6$ total dipion  widths of
$(5,n'), n'=1,2,3$ transition, and dipion spectra and $\cos
\theta$ distributions of ($5, n')$ transitions. We have succeeded
in explaining  approximately all 4 points, mentioned in
introduction:

\begin{enumerate}
    \item  Total widths $\Gamma_{\pi\pi}(5, n')$ are $O(1$ MeV).

    \item the sequence of inequalities between $\Gamma_{BB},
    \Gamma_{BB^*}, \Gamma_{B^*B^*}$ and corresponding widths for
    $B_sB_s$, occur naturally.
    \item Dikaon width of (5,1) is $\approx 1/10$ of the
    corresponding dipion width.
    \item Dipion spectra of (5,1), (5,2) transitions require
    inclusion of  FSI with $\sigma$ and $f_0$ peaks and the
    appearance of the peak at $M_{\pi\pi}\approx 0.4$ GeV is
    possible due to a nearby zero of amplitude. We stress, that our method allows to reproduce the sophisticated (5,1) spectrum in Fig.3
     with good accuracy, using the same FSI parameters as for the (5,2) spectrum in Fig. 5.
    \item  In addition the unusual ($U$-type) $\cos \theta$ dependence
    is quantitatively
    explained for the (5,2) transition as consequence of FSI.
\end{enumerate}

We have observed strong dependence of all results on the
properties of the $ \Upsilon (5S)$ wave function, in particular on
the position of its zeros, which in turn may serve to derive it
from the total set of experimental data.

As a whole, our method allows to understand the basic features of
all $\Upsilon (nS)$ transitions and decays, however more work is
needed to explain all data in detail.

The authors are grateful to M.V.Danilov and  S.I.Eidelman for
constant support and  suggestions, to P.N.Pakhlov and all members
of ITEP experimental group for stimulating discussions. The
financial support of  grants RFFI  06-02-17012, 06-02-17120 and
NSh-4961.2008.2 is gratefully acknowledged.
\begin{center}

\includegraphics[width=8cm,keepaspectratio=true]{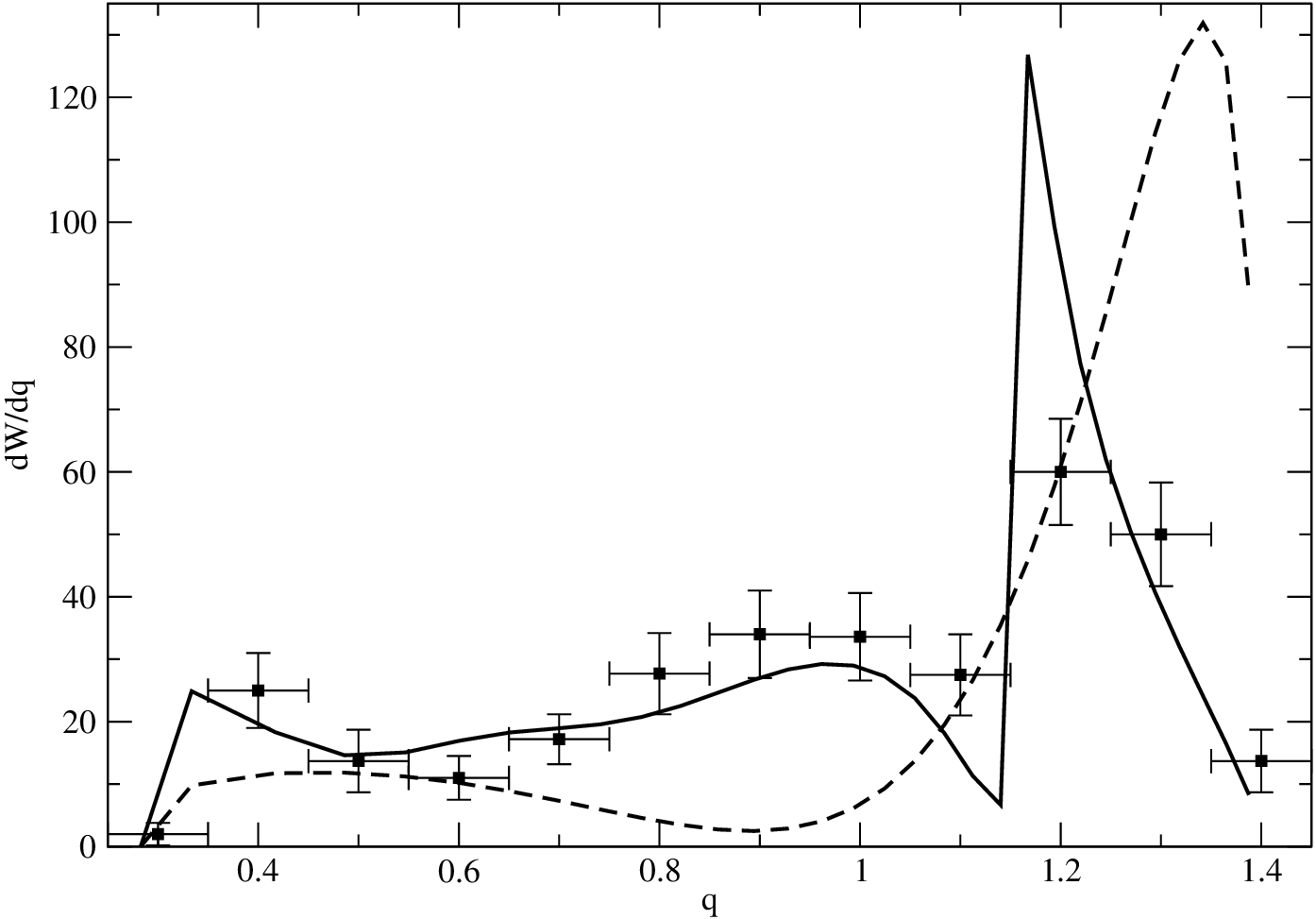}
\end{center}
Fig.3. Comparison of theoretical predictions, Eqs. (\ref{0}),
(\ref{00})  with experiment \cite{9} for the dipion spectrum,
 $\frac{dw}{dq}$, in the  $\Upsilon (5,1)\pi\pi$ transition. Theory: Eq. (\ref{0}) with
$f \equiv 1$ -- broken curve, Eq. (\ref{0}) with $f$
 as in Eq. (\ref{00}) (parameters given in the text) -- solid line. Theoretical curve
  is normalized to the total experimental width
  $\Gamma^{\exp}_{\pi\pi}=\frac{dw}{dq} dq$.

\begin{center}

\includegraphics[width=8cm,keepaspectratio=true]{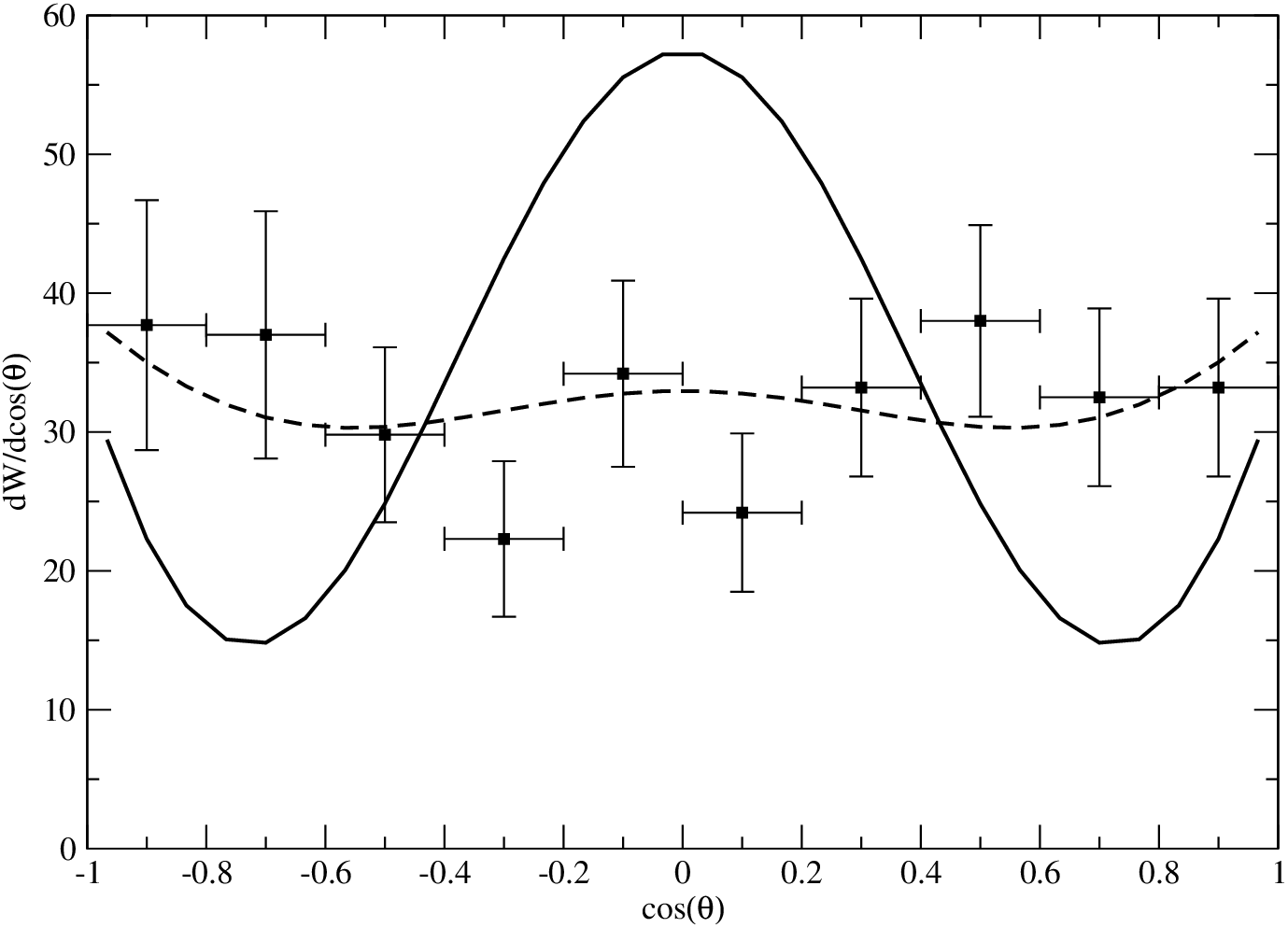}
\end{center}
Fig.4. The same as in Fig.3, for the angular distribution
$\frac{dw}{d\cos\theta}$ in the  $\Upsilon (5,1)\pi\pi$
transition.

\begin{center}
\includegraphics[width=8cm,keepaspectratio=true]{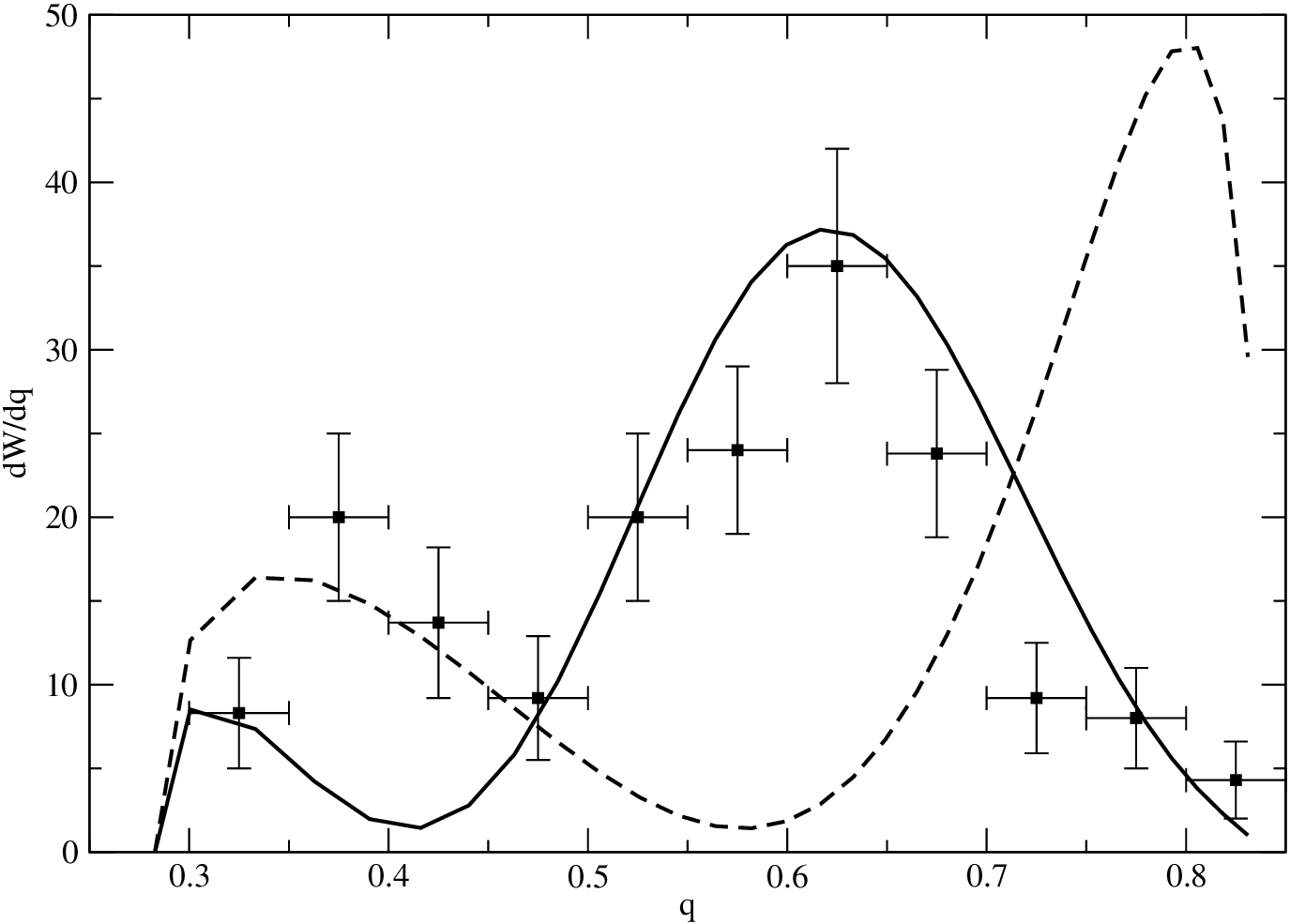}
\end{center}

Fig.5. The same as in Fig.3, for the dipion spectrum
$\frac{dw}{dq}$ in the
 $\Upsilon (5,2)\pi\pi$ transition.

\begin{center}
\includegraphics[width=8cm,keepaspectratio=true]{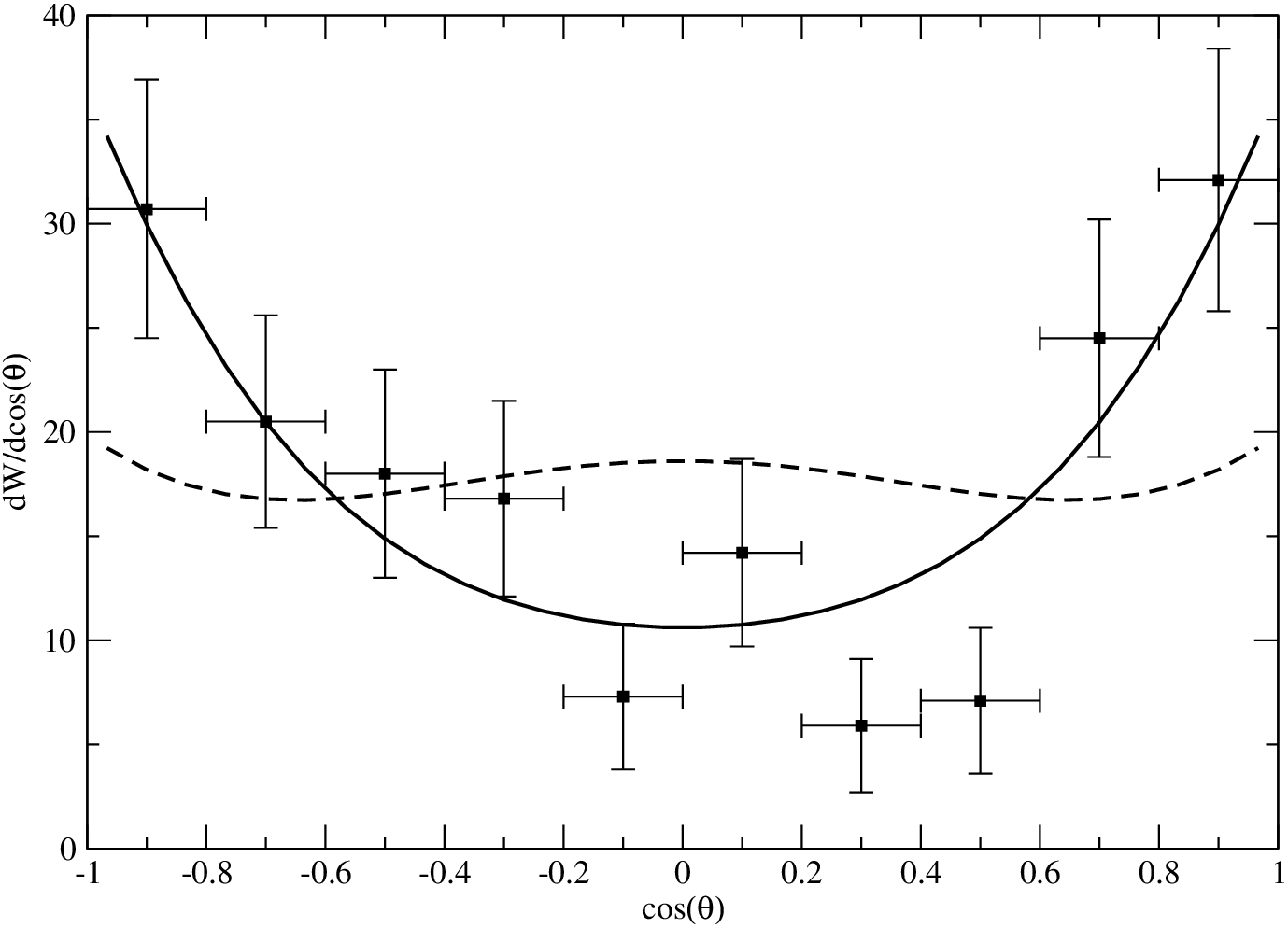}

\end{center}
Fig.6.  The same as in Fig.3, for the  angular distribution
$\frac{dw}{d\cos\theta}$   in the
 $\Upsilon (5,2)\pi\pi$ transition.


\newpage

\bigskip

{\bf Table 1.\\}{ The  values of  two-body decay widths $
\Gamma_k$  calculated with realistic $5S$ wave function.}

 \begin{center}
\vspace{3mm}

\begin{tabular}{|l|l|l|l|l|l|l|} \hline

&&&&&&\\
$k$& $1, B\bar B$& $2, B\bar B^*$& 3, $B^*\bar B^*$& $4, B_s\bar
B_s$ & $5, B_s\bar B_s^*$ & $6, B_s^* \bar B^*_s$\\ \hline

$p_k$, GeV & 1.26&1.16&1.05& 0.835& 0.683& 0.482\\\hline $M_k$,
GeV &5.28& 5.30&5.32&5.37&5.39&5.41\\\hline
 $Z_k$
&1&4&7&1/2&4/2&7/2\\\hline $\Gamma_k/\left( \frac{M_\omega}{2\omega}\right)^2$ MeV&11&57&65&0.08&10&18\\
\hline

\end{tabular}

\end{center}

{\bf Table 2.} The total dipion and dikaon widths for the models 1
and 2   (from top to bottom) in comparison with experimental
widths from \cite{9}.

 \begin{center}
\vspace{3mm}

\begin{tabular}{|l|l|l|l|l|} \hline

&&&&\\
transition&51&52&53&51,KK\\
$(n,n')$&&&&\\
\hline

$\Gamma^{AZI}_{\pi\pi}/\left(\frac{M_{br}}{f_\pi}\right)^4$,&
1.4&0.67&0.032&0.12\\MeV&&&&\\ \hline

$\Gamma^{FSI}_{\pi\pi}/\left(\frac{M_{br}}{f_\pi}\right)^4$,&
2.0&1.67&0.23&0.18\\MeV&&&&\\ \hline
$\Gamma_{\pi\pi}^{\exp}(n,n')$&$0.59\pm0.04$&$0.85\pm0.07$&$0.52^{+0.20}_{-0.17}$&$0.067^{+0.017}_{-0.015}$\\
MeV&$\pm 0.09$&$\pm 0.16$&$\pm 0.10$&$\pm 0.013$\\
\hline

\end{tabular}

\end{center}
\end{document}